\documentclass[twocolumn,showpacs,prc,superscriptaddress,amsmath,amssymb]{revtex4-1}

\usepackage{multirow}
\usepackage{graphicx}
\usepackage{array}
\usepackage{setspace}
\newcommand{\goodchi}{\protect\raisebox{2pt}{$\chi$}}
\begin{document}

\preprint{APS/123-QED}

\title{Entrance Channel Effects on the Quasifission Reaction Channel in Cr + W Systems}

\author{K. Hammerton}
\email{hammerto@nscl.msu.edu}
\affiliation{National Superconducting Cyclotron Laboratory, Michigan State University, East Lansing, Michigan 48824, USA}
\affiliation{Department of Chemistry, Michigan State University, East Lansing, Michigan 48824, USA}
\author{D. J. Morrissey}
\affiliation{National Superconducting Cyclotron Laboratory, Michigan State University, East Lansing, Michigan 48824, USA}
\affiliation{Department of Chemistry, Michigan State University, East Lansing, Michigan 48824, USA}
\author{Z. Kohley}
\affiliation{National Superconducting Cyclotron Laboratory, Michigan State University, East Lansing, Michigan 48824, USA}
\affiliation{Department of Chemistry, Michigan State University, East Lansing, Michigan 48824, USA}
\author{D.~J. Hinde}
\affiliation{Department of Nuclear Physics, Research School of Physics and Engineering, Australian National University, Canberra, Australian Capital Territory 2601, Australia}
\author{M. Dasgupta}
\affiliation{Department of Nuclear Physics, Research School of Physics and Engineering, Australian National University, Canberra, Australian Capital Territory 2601, Australia}
\author{A. Wakhle}
\affiliation{National Superconducting Cyclotron Laboratory, Michigan State University, East Lansing, Michigan 48824, USA}
\affiliation{Department of Nuclear Physics, Research School of Physics and Engineering, Australian National University, Canberra, Australian Capital Territory 2601, Australia}
\author{E. Williams}
\affiliation{Department of Nuclear Physics, Research School of Physics and Engineering, Australian National University, Canberra, Australian Capital Territory 2601, Australia}
\author{I.~P. Carter}
\affiliation{Department of Nuclear Physics, Research School of Physics and Engineering, Australian National University, Canberra, Australian Capital Territory 2601, Australia}
\author{K.~J. Cook}
\affiliation{Department of Nuclear Physics, Research School of Physics and Engineering, Australian National University, Canberra, Australian Capital Territory 2601, Australia}
\author{J. Greene}
\affiliation{Physics Division, Argonne National Laboratory, Lemont, Illinois 60473, USA}
\author{D.~Y. Jeung}
\affiliation{Department of Nuclear Physics, Research School of Physics and Engineering, Australian National University, Canberra, Australian Capital Territory 2601, Australia}
\author{D.~H. Luong}
\affiliation{Department of Nuclear Physics, Research School of Physics and Engineering, Australian National University, Canberra, Australian Capital Territory 2601, Australia}
\author{S.~D. McNeil}
\affiliation{Department of Nuclear Physics, Research School of Physics and Engineering, Australian National University, Canberra, Australian Capital Territory 2601, Australia}
\author{C. Palshetkar}
\affiliation{Department of Nuclear Physics, Research School of Physics and Engineering, Australian National University, Canberra, Australian Capital Territory 2601, Australia}
\author{D.~C. Rafferty}
\affiliation{Department of Nuclear Physics, Research School of Physics and Engineering, Australian National University, Canberra, Australian Capital Territory 2601, Australia}
\author{C. Simenel}
\affiliation{Department of Nuclear Physics, Research School of Physics and Engineering, Australian National University, Canberra, Australian Capital Territory 2601, Australia}
\author{K. Stiefel}
\affiliation{National Superconducting Cyclotron Laboratory, Michigan State University, East Lansing, Michigan 48824, USA}
\affiliation{Department of Chemistry, Michigan State University, East Lansing, Michigan 48824, USA}

\date{\today}

\begin{abstract}
\noindent
\textbf{Background:}
Formation of a fully equilibrated compound nucleus is a critical step in the heavy-ion fusion reaction mechanism but can be hindered by orders of magnitude by quasifission, a process in which the dinuclear system breaks apart prior to full equilibration. To provide a complete description of heavy-ion fusion it is important to characterize the quasifission process. In particular, the impact of changing the neutron-richness of the quasifission process is not well known. A previous study of Cr~+~W reactions at a constant 13 $\%$ above the Coulomb barrier concluded that an increase in neutron-richness leads to a decrease in the prominence of the quasifission reaction channel. \\
\textbf{Purpose:}
The interplay between the fusion-fission and quasifission reaction channels, with varying neutron-richness, was explored at a constant excitation energy, closer to the interaction barrier than the previous work, to see if the correlation between neutron-richness and quasifission is valid at lower energies.\\
\textbf{Methods:}
Mass distributions were determined for eight different combinations of Cr~+~W reactions at the Australian National University at 52.0 MeV of excitation energy in the compound nucleus.\\
\textbf{Results:}
A curvature parameter was determined for the fission-like fragment mass distributions and compared to various reaction parameters known to influence quasifission.\\
\textbf{Conclusions:}
The present work demonstrates that at energies near the interaction barrier the deformation effects dominate over the neutron-richness effects in the competition between quasifission and compound nucleus formation in these Cr~+~W systems and is an important consideration for future with heavy and superheavy element production reactions.

\end{abstract}


\maketitle

\section{Introduction}
\par
The fusion of two large nuclei has thus far been the primary mechanism for the formation of superheavy nuclei~\cite{Oganessian2006, Oganessian2012, Oganessian2007,Hamilton2013}.  There is great interest in producing new superheavy nuclei because each additional nucleon furthers our understanding of the limits of nuclear stability~\cite{Hamilton2013,Hofmann2000} and there are predictions that there will be a spherical shell closure near N = 184 and Z $\approx$ 114-126~\cite{Bender1999}.  Experimental work has already shown indications of a region of enhanced stability in neutron-rich nuclei near Z~$\geq 110$ and N~$\approx$~171 - 174~\cite{Oganessian2010, Hamilton2013}. However, even these very neutron-rich nuclei are still $\sim$ 10 neutrons away from N=184.  To reach nuclei in the N~=~184 region more neutron-rich projectiles and targets than the commonly used stable $^{48}$Ca and actinide targets will be necessary ~\cite{Hamilton2013,Oganessian2012,Bender1999,Bender2001,Nazarewicz2002}.  The next generation rare isotope facilities will allow exploration of the heavy ion fusion mechanism with medium mass neutron-rich projectiles that can form neutron-rich low mass superheavy nuclei~\cite{Hofmann2001, Loveland2007, Smolaczuk2010}.  Therefore, it is vital to have an understanding of the effect of increasing the neutron-richness of the system on the heavy-ion fusion reaction mechanism.

The cross section for the formation of a superheavy evaporation residue $\sigma_{evr}$ has been written as
\begin{equation}
\sigma_{evr} = \sum_{\mathrm{J}=0}^{\mathrm{J_{max}}} \sigma_{\mathrm{cap}}(\mathrm{J})\mathrm{P_{CN}(\mathrm{J})W_{sur}}(\mathrm{J}),
\end{equation}
where J is the angular momentum, $\mathrm{\sigma_{cap}}$ is the capture cross section for a given entrance channel, P$\mathrm{_{CN}}$ is the probability of forming a compound nucleus, and W$_{\mathrm{sur}}$ is the probability of the compound nucleus surviving against fission~(\cite{Hinde2002a} and references therein).  Following capture, formation of a fully fused compound nucleus can be hindered by the early separation of the dinuclear system, termed quasifission~\cite{Back1985,Toke1985}.  Quasifission has been shown to hinder fusion by orders of magnitude~\cite{Berriman2001, Yanez2013, Hamilton2013} in some cases.  A large effort has focused on understanding the entrance channel conditions that favor quasifission including: mass asymmetry~\cite{Berriman2001}, fissility of the compound nucleus~\cite{Swiatecki1982,Bjornholm1982}, reaction energy~\cite{Tsang1983, Back1985}, magicity~\cite{Itkis2007,Simenel2012}, and neutron-richess of the compound nucleus (N/Z)$_{\mathrm{CN}}$~\cite{Simenel2012,Sahm1985,Lesko1986, Vinodkumar2008,Liang2012,Adamian2000,Wang2010, Hammerton2015}.  Heavy-ion fusion is further complicated by entrance channel nuclear structure effects including large static deformations in the heavy reaction partner~\cite{Hinde1995,Nishio2001,Lin2012, Wakhle2014}.

So called mass-angle distributions (MAD) have been used extensively to study quasifission reaction dynamics~\cite{Hinde2008a, Wakhle2014, DuRietz2013}.  Ref~\cite{DuRietz2013} provides an overall view of MADs from reactions of medium mass projectiles and targets.  Three regions were identified based on the shape of the mass distribution and the entrance channel charge product Z$_{\mathrm{p}}$Z$_{\mathrm{t}}$.  The study of reactions at the intersection of two of these regions will provide important information on the quasifission mechanism. Cr + W is a prime candidate to study as its charge product 1776 is at the intersection of reactions that show short time-scale quasifission (the dinuclear system separated after very little rotation with partial mass transfer) and medium time scale reactions (the system rotated through larger angles and the fragments have time to move towards mass equilibration)~\cite{DuRietz2013}.

Previous measurements of the Cr + W systems, at beam energies chosen to give a constant ratio to the respective interaction barriers V$_{\mathrm{B}}$~\cite{Bass1977a} ,showed that the compound nucleus N/Z$_{\mathrm{CN}}$ was important in determining the relative amount of quasifission~\cite{Hammerton2015}.  In the present work, the effect of changing the neutron-richness of the compound nucleus was explored for the same Cr + W systems at a constant excitation energy E$^{*}=52.0$ MeV, closer to the interaction barrier than the reactions reported in~\cite{Hammerton2015} and similar to that used in hot-fusion reactions.

\par

\section{Experimental Details}

\begin{table*}
\caption{Entrance channel, compound nucleus formed, number of neutrons relative to the compound nucleus $^{230}$ formed by $^{50}$Cr $+ ^{180}$W, W target thickness, center of mass energy E$_{\mathrm{c.m.}}$, excitation energy E$^{*}_{\mathrm{CN}}$, E$_{\mathrm{c.m.}} / \mathrm{V}_{\mathrm{B}}$, $\mathrm{(N/Z)}_{\mathrm{CN}}$, calculated $l_{\mathrm{max}}$ from the total reaction cross section~\cite{Tarasov2003}, and $l_{\mathrm{crit}}$~\cite{Tarasov2003} for each of the systems.}
 \centering
 \setlength{\tabcolsep}{5.5pt}
 \setlength{\extrarowheight}{2pt}
\begin{tabular}{ccccccccc}
  \hline
  \hline
 System & $\Delta$N & \begin{tabular}{@{}c@{}}Target Thickness \\ ($\mu$g cm$^{-1}$)\end{tabular}& \begin{tabular}{@{}c@{}}E$_{\mathrm{c.m.}}$ \\ (MeV) \end{tabular} & \begin{tabular}{@{}c@{}}E$^{*}_{\mathrm{CN}}$ \\ (MeV) \end{tabular} & E$_{\mathrm{c.m.}} / \mathrm{V}_{\mathrm{B}}$ & $\mathrm{(N/Z)}_{\mathrm{CN}}$ & \begin{tabular}{@{}c@{}} $l_{\mathrm{max}}$ \\ ($\hbar$)\end{tabular} & \begin{tabular}{@{}c@{}} $l_{\mathrm{crit}}$ \\ ($\hbar$)\end{tabular} \\
  \hline
  $^{50}$Cr $+ ^{180}$W$\rightarrow ^{230}$Cf & 0& 48 & 210.0&52.0&1.07&1.35& 67 & 58 \\
  $^{50}$Cr $+ ^{186}$W$\rightarrow ^{236}$Cf & 6& 43 & 201.3&52.0&1.03&1.41& 44 & 39 \\
  $^{52}$Cr $+ ^{180}$W$\rightarrow ^{232}$Cf & 2& 48 & 214.1&52.0&1.09&1.37& 76 & 71 \\
  $^{52}$Cr $+ ^{184}$W$\rightarrow ^{236}$Cf & 6& 64 & 209.7&52.0&1.08&1.41& 72 & 64 \\
  $^{54}$Cr $+ ^{180}$W$\rightarrow ^{234}$Cf & 4& 46 & 215.4&52.3&1.11&1.39& 85 & 76 \\
  $^{54}$Cr $+ ^{182}$W$\rightarrow ^{236}$Cf & 6& 97 & 213.8&52.0&1.10&1.41& 81 & 75 \\
  $^{54}$Cr $+ ^{184}$W$\rightarrow ^{238}$Cf & 8& 64 & 211.8&52.0&1.09&1.43& 76 & 72 \\
  $^{54}$Cr $+ ^{186}$W$\rightarrow ^{240}$Cf & 10& 43 & 209.5&52.0&1.08&1.45& 72 & 69 \\
  \hline
  \hline
\end{tabular}
\label{tab:tableI}
\end{table*}

\par
\begin{figure}
\includegraphics*[width=0.45\textwidth]{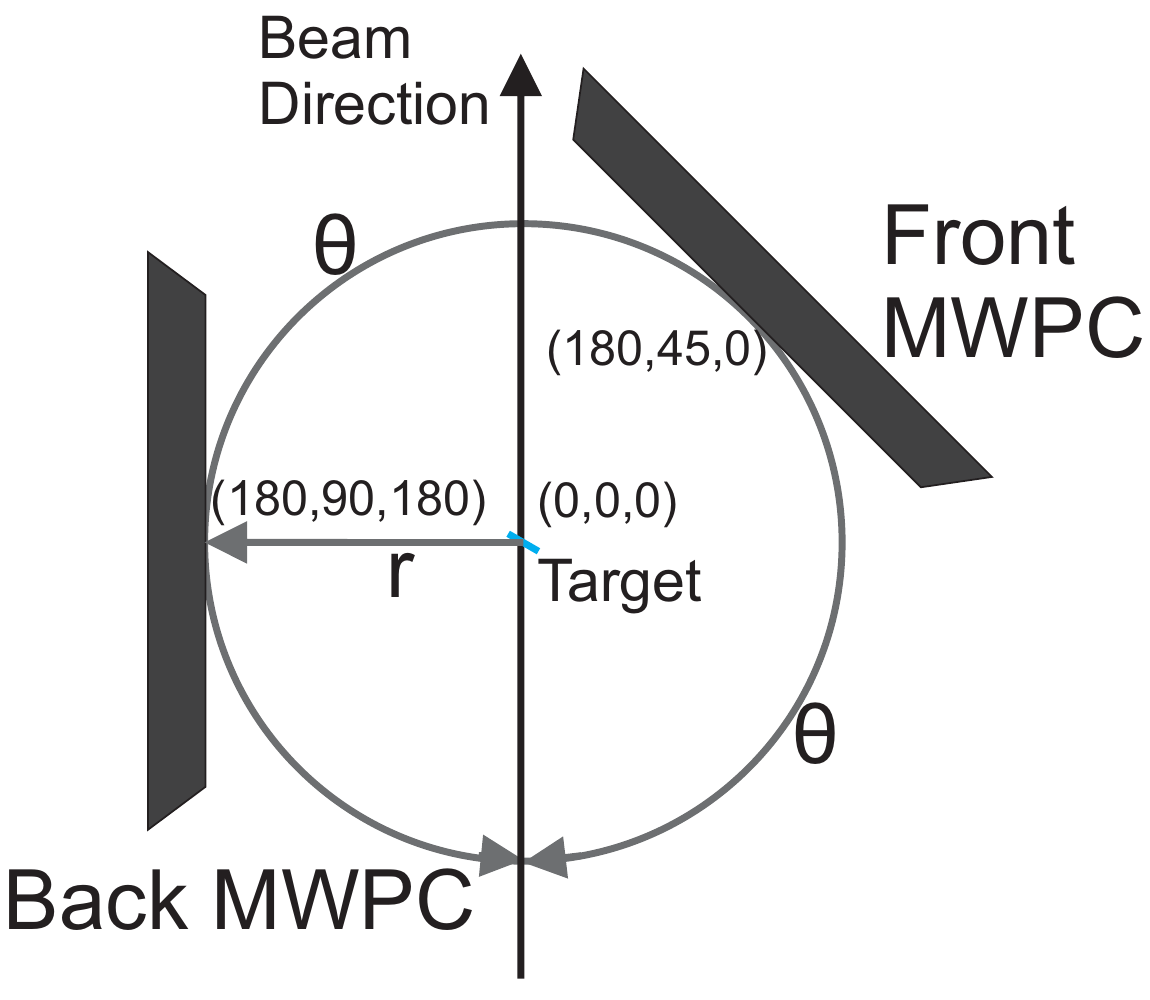}
\caption{Schematic scale diagram of the CUBE detector setup from above. The definitions of $\theta$ and r are indicated. The coordinates (r,$\theta$,$\Phi$) at the center of the CUBE and at the center of the two MWPCs are indicated, in mm and degrees.}
\label{fig:cube}
\end{figure}

Beams of $^{50,52,54}$Cr provided by the 14UD electrostatic accelerator and superconducting LINAC at the Heavy Ion Accelerator Facility at the Australian National University (ANU) were used to bombard isotopically enriched targets of $^{180,182,184,186}$W with thicknesses ranging from $43-97~\mu$g/cm$^{2}$ mounted on $40-60~\mu$g/cm$^{2}$ carbon backings~\cite{Greene2015}. The details are given in Table~\ref{tab:tableI}. Fragments resulting from fusion-fission and quasifission reactions (collectively termed fission-like) were detected in coincidence using the ANU CUBE detector system~\cite{Hinde1996}. The detector system consisted of two large-area, position-sensitive multiwire proportional counters (MWPCs). A diagram of the CUBE detector set-up used in the present work is shown in Figure~\ref{fig:cube}. Each MWPC had an active area of $28\times36$ cm$^2$~\cite{Hinde1996, DuRietz2013} that were used to cover laboratory scattering angles of $5^{\circ} < \theta < 80^{\circ}$ and $ 50^{\circ} < \theta < 125^{\circ}$. Time-of-flight and position information for coincident fission fragment events were used to calculate the mass distribution over all measured angles for full momentum transfer events using the kinematic coincidence technique~\cite{Hinde2008,DuRietz2013}.
From the measured velocity vectors of the coincident fragments the mass ratio, $M_{\mathrm{R}} = m_{1} / (m_{1} + m_{2})$ where $m_1$ and $m_2$ are the masses of the fission fragments, could be determined~\cite{DuRietz2013}.

\par
From the list of the measured reactions given in Table~\ref{tab:tableI} one can see that the most neutron-deficient ($^{50}$Cr $+ ^{180}$W) and neutron-rich ($^{54} $Cr $+ ^{186}$W) systems are different by ten neutrons, which provided an opportunity to study the (N/Z)$_{\mathrm{CN}}$ dependence of the reactions while holding other variables that are known to affect the quasifission process constant.  The Cr~+~W reactions all have the same entrance channel charge product, Z$_{\mathrm{p}}$Z$_{\mathrm{t}}$, of 1776 and only the $^{52}$Cr has a closed shell at $N = 28$. The W targets are deformed and have calculated $\beta_{2}$ values in the range of 0.225 to 0.254 ~\cite{Moller1995}.

\section{Results and Discussion}

\begin{figure*}
\includegraphics*[width=0.9\textwidth]{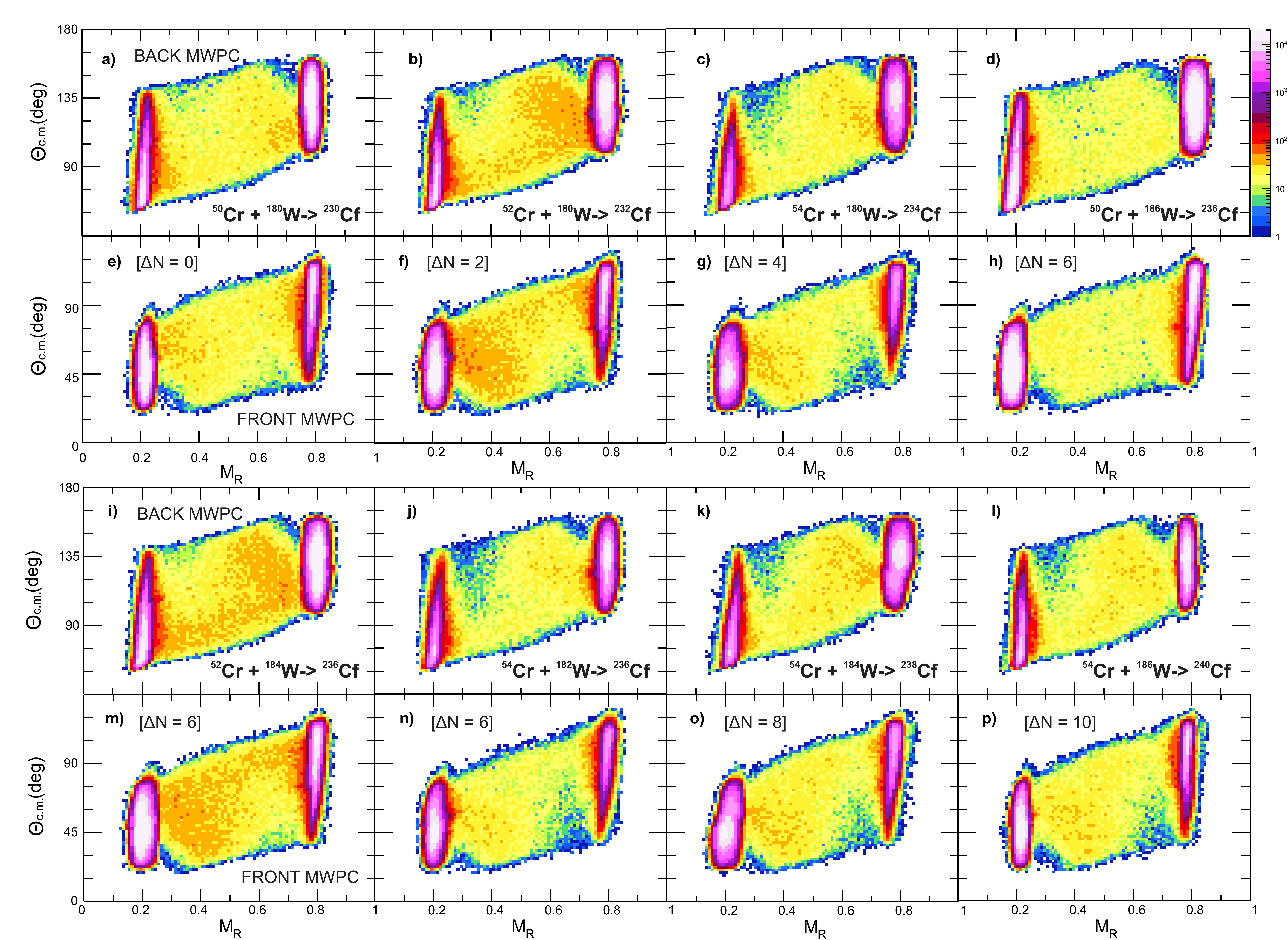}
\caption{(color online) The mass-angle distributions are shown for all eight Cr + W systems from the present work at E$^{*} = 52.0$ MeV. In the first (Panels a-d) and third rows (Panels i-l) the MADs from the Back MWPC are shown.  In the second (Panels e-h) and fourth (Panels m-p) rows the MADs from the Front MWPC are shown. For each system, the projectile, target, and number of neutrons relative to $^{230}$Cf is given.}
\label{fig:MADists}
\end{figure*}

The effect of changing the neutron-richness on the reaction dynamics was explored by analyzing the mass-angle distributions (MAD) generated from the deduced mass ratios and center-of-mass angles ($\theta_{\mathrm{c.m.}}$)~\cite{Thomas2008}. The MADs for all eight of the Cr~+~W systems are shown in Figure~\ref{fig:MADists}. The MADs observed in the Back MWPC are shown in Panels a-c and i-l, and the MADs observed in the the Front MWPC are shown in Panels e-h and m-p. The intense bands of events at M$_{\mathrm{R}} \sim$ 0.2 and 0.8 result from elastic scattering events. The region between these two bands contains events from quasifission and fusion-fission, termed the fission-like region. In the present work, the fission-like region was defined to be between mass ratios of 0.35 and 0.65. A notable feature in the fission-like region is the correlation between mass and angle. This correlation has been interpreted in the literature as a signature of quasifission.  At center-of-mass energies greater than $\sim 20$ MeV, fragments from fusion-fission reactions will be found at all angles and will form a narrow peak in the mass ratio distribution at M$_{\mathrm{R}} = 0.5$~\cite{Colby1960, Toke1985}.  A correlation between mass and angle was observed in each MAD in Figure~\ref{fig:MADists} indicating that in addition to fusion-fission a quasifission component is present.

\begin{figure*}
\includegraphics*[width=0.9\textwidth]{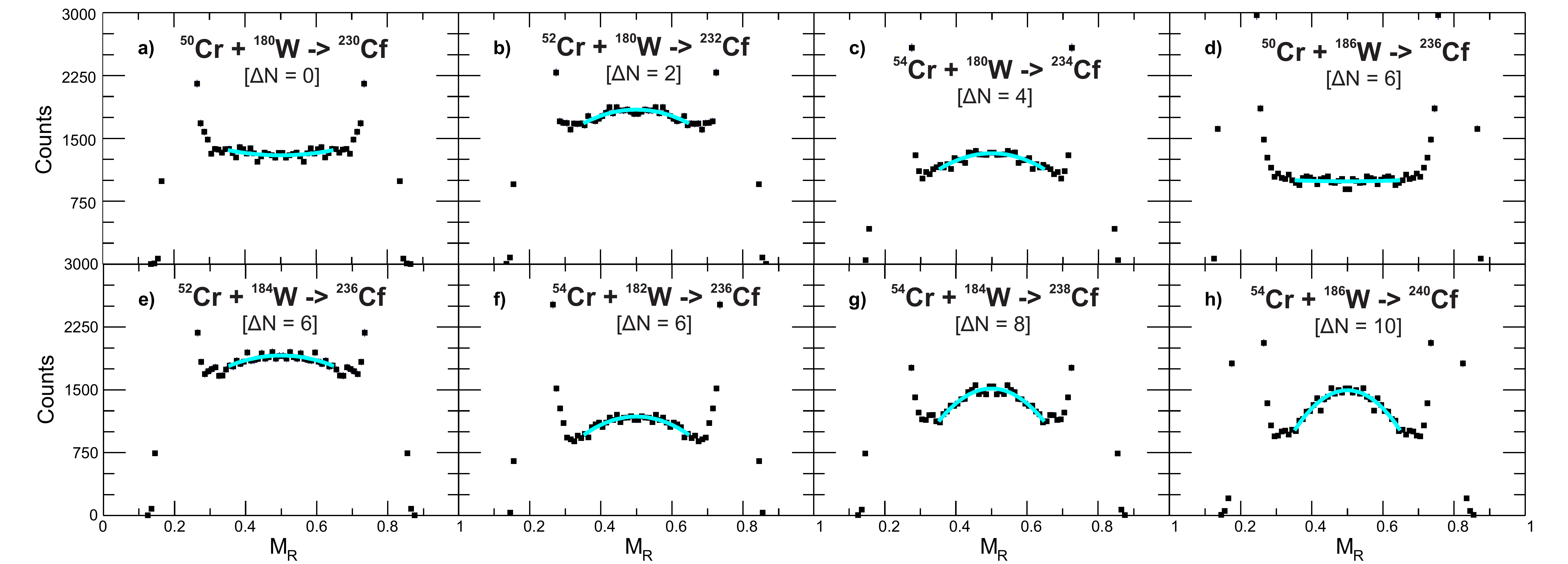}
\caption{(color online) The projected mass distributions are shown for all eight Cr + W systems from the present work at E$^{*} = 52.0$ MeV.  The solid (black) circles represent the experimental data points.  Experimental uncertainties are smaller than the size of the points.  The solid (blue) line shows a quadratic fit to the fission-like region. For each systems, the projectile, target, compound nucleus, and number of neutrons relative to $^{230}$Cf is given.}
\label{fig:MassDists}
\end{figure*}

The angle integrated mass distributions generated for all eight of the Cr + W systems are shown in Figure~\ref{fig:MassDists}. The large, sharp peaks at M$_{\mathrm{R}} \sim$ 0.2 and 0.8 result from elastic scattering events.  In the literature a broadening of the mass distribution in the fission-like region has been interpreted  as an increase in the prominence of the quasifission reaction channel~\cite{Lebrun1979}.

Previously~\cite{Hinde1996, Prokhorova2008, Williams2013, Hinde2008, Itkis2011, Itkis2004, Lin2012, Hammerton2015, Simenel2012} mass distributions were compared by fitting the fission-like region with a Gaussian function and extracting the width of the resulting Gaussian function. However, this method was not appropriate in the present work. In Figure~\ref{fig:MassDists}, two systems, $^{50}$Cr$ + ^{180}$W (\ref{fig:MassDists}~a) and $^{50}$Cr$ + ^{186}$W (\ref{fig:MassDists}~d), stand in contrast to the others. Rather than having a maximum at a mass ratio of M$_{\mathrm{R}} = 0.5$, these two distributions have a minimum and cannot be characterized with a Gaussian function.  Instead, each mass distribution was fitted in the fission-like region with a second degree polynomial function. The results of these fits are represented by the solid (blue) lines in each panel of Figure~\ref{fig:MassDists}. The second derivative, determined as two times the second order coefficient of the function resulting from the fit, was used as a quantitative measure of the curvature of the mass distributions and called the curvature parameter. A more negative curvature parameter indicates that the mass distribution has a narrower peak in the fission-like region and the quasifission reaction channel is less prominent than in a system with a larger curvature parameter.

\begin{figure}
\centering
\includegraphics*[width=.5\textwidth]{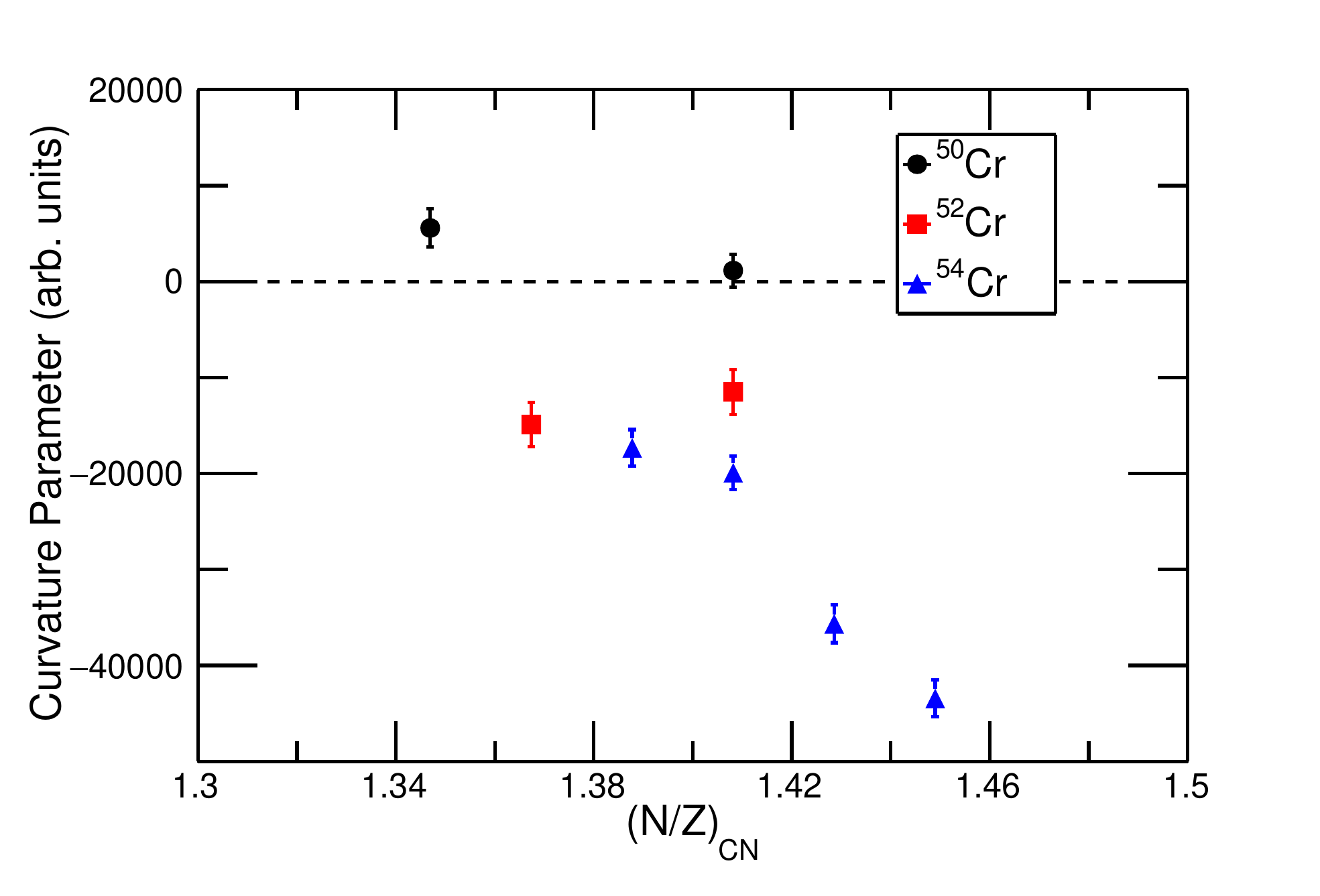}
\caption{The curvature parameters determined from the fit of the fission-like region of the mass distributions are shown as a function of (N/Z)$_{\mathrm{CN}}$.}
\label{fig:CurvVsNZee}
\end{figure}

The curvature parameters determined for the mass distributions of the Cr~+~W systems are shown as a function of (N/Z)$_{\mathrm{CN}}$ in Figure~\ref{fig:CurvVsNZee}. While there is a general decrease in curvature with increasing neutron-richness the correlation is not as notable as that previously reported for the Cr~+~W systems measured at 13$\%$ above the Bass barrier~\cite{Bass1977a} reported in~\cite{Hammerton2015}. Two systems of note are the two where $^{50}$Cr was the projectile which have a positive curvature indicating that the quasifission reaction channel was particularly enhanced (or the quasifission time scale was shorter) in these two systems compared to the other systems.

\subsection{Bohr Independence Hypothesis}

The Bohr Independence Hypothesis~\cite{Bohr1936} states that once a nucleus, with a given angular momentum, fully equilibrates in all degrees of freedom it loses all memory of the entrance channel. Thus, the decay of an equilibrated compound nucleus should be independent of the entrance channel through which it was produced.

The mass distributions of three systems where $^{236}$Cf was the compound nucleus are shown in Panels d-f of Figure~\ref{fig:MassDists}.  Note that in all three of these systems the compound nucleus was formed at the same excitation energy, 52.0 MeV.  If compound nucleus formation was the dominant reaction channel in these systems then the Bohr Independence Hypothesis should apply and the decay of these three systems should be the same.  The three mass distributions, however, are not the same.  The mass distribution from the reaction of $^{54}$Cr$ + ^{182}$W$\rightarrow ^{236}$Cf has a narrower mass distribution than that of $^{50}$Cr$ + ^{186}$W$\rightarrow ^{236}$Cf and a curvature parameter that is a factor of $\sim$17 smaller.  The mass distribution of $^{52}$Cr$ + ^{184}$W$\rightarrow ^{236}$Cf falls in between the other two systems. This difference among the systems can be observed in Figure~\ref{fig:CurvVsNZee} where the curvature parameters determined for these systems are shown as a function of (N/Z)$_{\mathrm{CN}}$ (1.41 for $^{236}$Cf). This is a deviation from the Bohr Independence Hypothesis and supports the conclusion that a significant fraction of the events in each of these mass distributions does not result from the decay of a fully fused compound nucleus, but rather undergoes an earlier separation via the quasifission reaction channel.

The possibility of different angular momentum distributions between each reaction system should be considered.  The $^{50}$Cr reactions, with flat mass distributions, have the lowest angular momenta (Table~\ref{tab:tableI}). The $^{54}$Cr reactions, with the narrowest mass distributions, have the highest angular momenta.  This is opposite to expectations if angular momentum alone was responsible to the increased mass widths~\cite{Lebrun1979}, thus the observation cannot be attributed to the angular momentum differences.  Other possible correlations are discussed below.

\subsection{Fissility and Mass Asymmetry}
The fissility of the compound nucleus $\goodchi_{\mathrm{CN}}$ and the mass asymmetry of the entrance channel $\alpha$ are two parameters that change with the entrance channel that are correlated with neutron-richness.  $\goodchi_{\mathrm{CN}}$ is inversely correlated with neutron-richness and defined as
\begin{equation}
\goodchi_{\mathrm{CN}}= (Z^{2}/A) / (Z^{2}/A)_{\mathrm{crit}}
\end{equation}
where
\begin{equation}
(Z^{2}/A)_{\mathrm{crit}} = 50.883(1-1.7826~I^{2})
\end{equation}
and
\begin{equation}
I = (A-2Z)/A
\end{equation}
~\cite{Swiatecki1982, Bjornholm1982}. Previous studies~\cite{Swiatecki1982, Bjornholm1982} of the $\goodchi_{\mathrm{CN}}$ dependence of quasifission observed a decrease in quasifission with decreasing $\goodchi_{\mathrm{CN}}$.

Mass asymmetry, defined as $\alpha = (A_{\mathrm{Target}} - A_{\mathrm{Proj}}) / (A_{\mathrm{Target}} + A_{\mathrm{Proj}})$ also decreases as the neutron-richness of the projectile increases.  Decreasing mass asymmetry, however, has been shown to lead to an increase in quasifission ~\cite{Berriman2001}.

\begin{figure*}
\centering
\includegraphics*[width=0.75\textwidth]{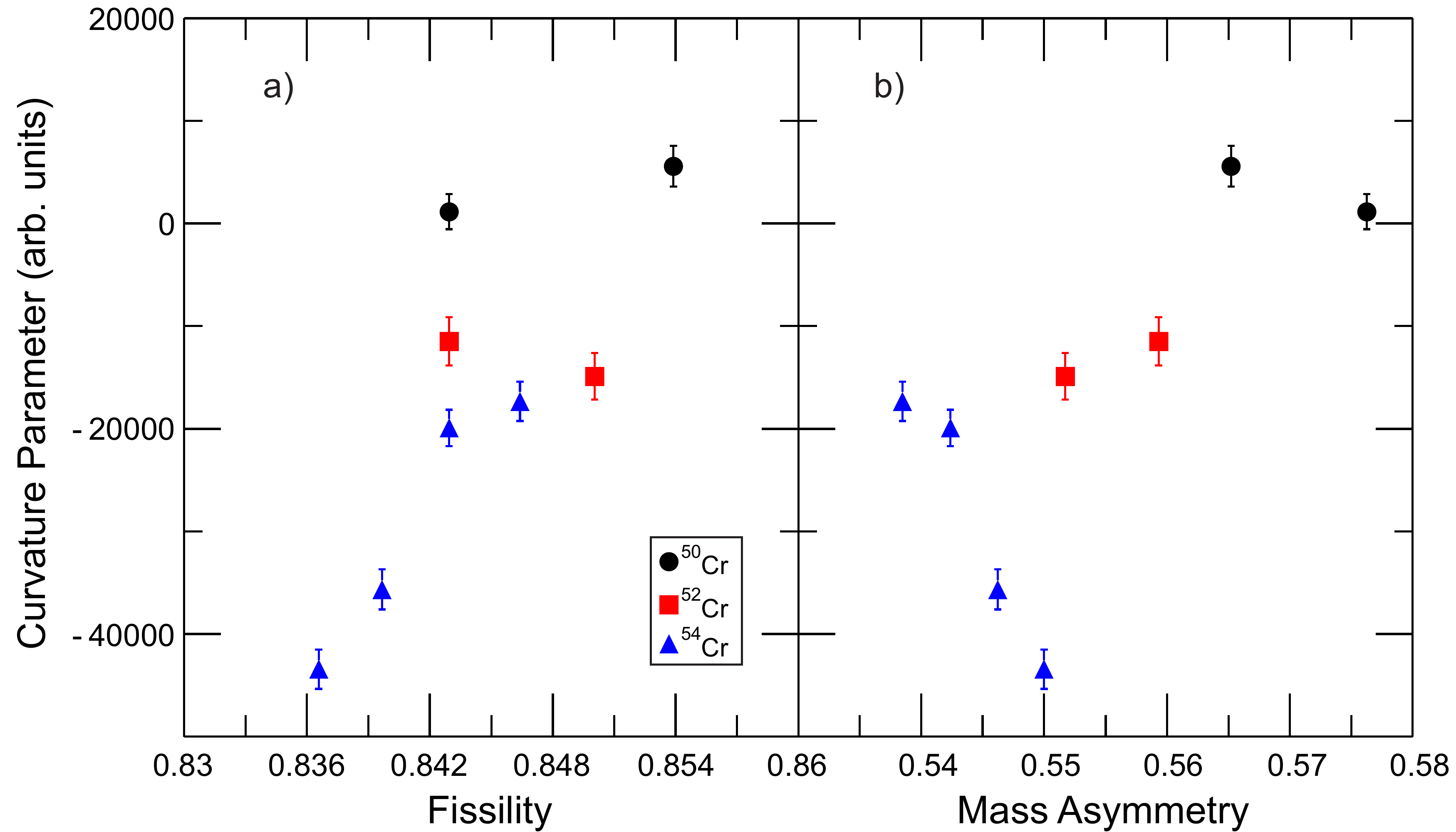}
\caption{The curvature parameter determined for each system as a function of fissility ($\goodchi_{\mathrm{CN}}$) (Panel a) and mass asymmetry ($\alpha$) (Panel b).}
\label{fig:fissAlpha}
\end{figure*}

In many common theoretical models~\cite{Zagrebaev2001a,Swiatecki1982,Bjornholm1982,Adamian2000,Back2014,Adamian2012, Antonenko1993,Antonenko1995,Blocki1986,Adamian2003} either the fissility or mass asymmetry is used as the dominant predictor of the importance of quasifission. Thus, the choice of a model has resulted in conflicting conclusions as to the nature of the influence of neutron-richness on quasifission because of the differences between the correlation with fissility and the correlation with mass asymmetry. For example, measurements by Lesko \textit{et al.}~\cite{Lesko1986} and Liang \textit{et al}.~\cite{Liang2012} of Sn~+~Ni systems showed that (N/Z)$_{\mathrm{CN}}$ increased as the quasifission flux increased.  However, measurements of Sn~+~Zr by Vinodkumar \textit{et al}.~\cite{Vinodkumar2008} and Sahm \textit{et al}.~\cite{Sahm1985} found a decrease in quasifission as (N/Z)$_{\mathrm{CN}}$ increased.   Observations from previously reported measurements of Cr + W reactions at E$_{\mathrm{c.m.}}/$V$_{\mathrm{B}}=1.13$~\cite{Hammerton2015} indicated that the collisions did not exclusively form compound nuclei due to the change in compound nuclear fissility as the neutron-richness increased. In Figure~\ref{fig:fissAlpha}, the curvature parameters determined for each system are shown as a function of $\goodchi_{\mathrm{CN}}$ and $\alpha$. There is considerable scatter of the curvature parameters when plotted against either $\goodchi_{\mathrm{CN}}$ or $\alpha$ although there is a general increase in the curvature parameter with increasing fissility.

\subsection{Energetics}

\begin{figure}
\centering
\includegraphics*[width=0.5\textwidth]{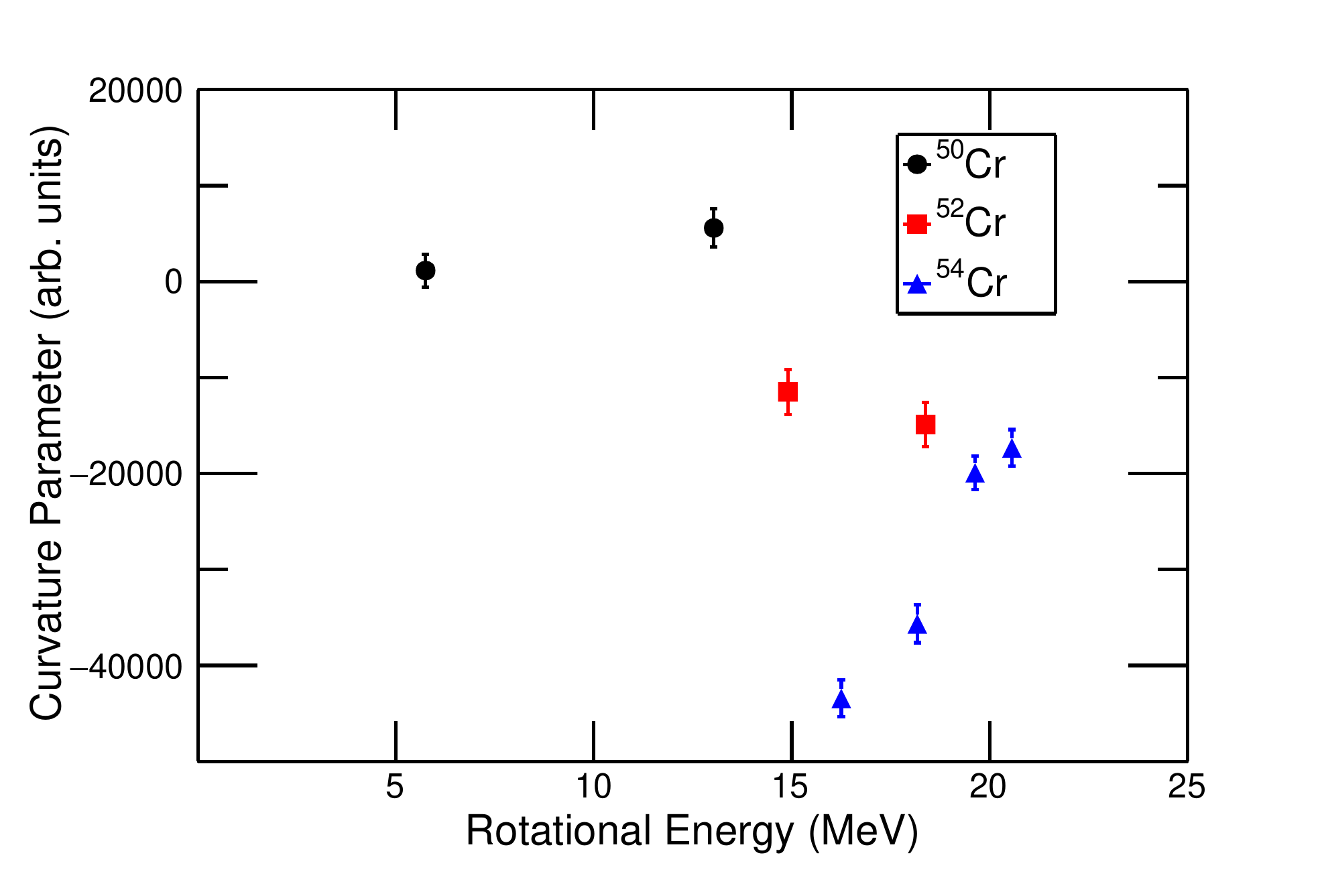}
\caption{(color online) The curvature parameter determined for each Cr~+~W system as a function of the maximum rotational energy E$_{\mathrm{rot}}$ (MeV).}
\label{fig:CurvVsErotee}
\end{figure}

The center of mass energies for each system differed significantly to reach the same E$^*_{\mathrm{CN}} = 52.0$ MeV.  This resulted in large variations in the maximum rotational energy available to each system, calculated as  E$_{\mathrm{rot}}$ = E$_{\mathrm{c.m.}}$ - V$_{\mathrm{B}}$.  The maximum rotational energies carried by for the systems measured in this work ranged from 5.75 to 20.56 MeV. While the systems with the lowest rotational energies have the highest curvature, there is not an overall correlation between maximum rotational energies and curvature, as seen in Figure~\ref{fig:CurvVsErotee}. Note that the previously reported Cr~+~W systems~\cite{Hammerton2015}, measured at E$_{\mathrm{c.m.}}$ /V$_{\mathrm{B}} = 1.13$, had a minimal change in rotational energy from 25.12 - 25.6 MeV, thus limiting the influence of this variable.

\subsection{Deformation Effects}

The $^{50,52,54}$Cr nuclei can be approximated as spherical, but $^{180,182,184,186}$W are all strongly deformed. The $\beta_{2}$ values calculated in~\cite{Moller1995} are shown in Table~\ref{tab:radii}. The significant deformation of the W nuclei results in a $\sim1$ fm variation from the average radius of the semi-major and semi-minor radius, Table~\ref{tab:deformation}.

\begin{table}
  \centering
  \caption{Bass average radii, $\beta_{2}$ values, semi-major radii, and semi-minor radii.}
  \setlength{\tabcolsep}{5pt}
 \setlength{\extrarowheight}{2pt}
\begin{tabular}{ccccc}
\hline \hline
Nucleus & \begin{tabular}{@{}c@{}}R(fm) \\ (average)\end{tabular} & \begin{tabular}{@{}c@{}}$\beta_{2}$\\ \cite{Moller1995}\end{tabular} &\begin{tabular}{@{}c@{}}R(fm) \\ (semi-major)\end{tabular}& \begin{tabular}{@{}c@{}}R(fm) \\ (semi-minor)\end{tabular}\\
  \hline
  $^{50}$Cr&3.89&0.0&--&--\\
  $^{52}$Cr&3.96&0.0&--&--\\
  $^{54}$Cr&4.02&0.0&--&--\\
  $^{180}$W&6.30&0.258&7.33&5.79\\
  $^{182}$W&6.33&0.259&7.36&5.81\\
  $^{184}$W&6.35&0.24&7.32&5.87\\
  $^{186}$W&6.38&0.23&7.30&5.92\\
  \hline
  \hline
\end{tabular}
  \label{tab:radii}
\end{table}

To reach E$^{*}_{\mathrm{CN}}$ of 52.0 MeV, the center of mass energy of the reaction is lower and thus closer to the Bass barrier~\cite{Bass1977a} relative to the energies necessary to reach E$_{\mathrm{c.m.}}/\mathrm{V}_{\mathrm{B}} = 1.13$.  As shown by previous works~\cite{Hinde1995, Hinde1996, Nishio2010, Wakhle2014, Knyazheva2007, Hinde2008,Nishio2008}, deformation has a large impact near the interaction barrier.  In reactions of deformed nuclei, the barrier is dependent on the orientation of the deformed nucleus.  Generally, the reported interaction barrier is an average of all possible collision orientations.

Many previous works~\cite{Hinde1995, Hinde1996, Hinde2002, Rafiei2008,Thomas2008,Hinde2008,Nishio2001,Mitsuoka2002} have shown that at E$_{\mathrm{c.m.}}$ near or below the interaction barrier the structure of the nuclei involved in a heavy-ion fusion reaction, particularly a heavy reaction partner with a large deformation, has a significant effect on the reaction dynamics.  When a deformed heavy nucleus takes part in the reaction the evaporation residue cross section was observed to be hindered at energies near the barrier~\cite{Nishio2001,Mitsuoka2002}. Similarly, at center-of-mass energies near the barrier, hindrance of the related fusion-fission reaction channel has also been attributed to the presence of a deformed heavy nucleus in the entrance channel ~\cite{Hinde1995, Hinde1996, Hinde2002, Rafiei2008,Thomas2008,Hinde2008}.  This loss in evaporation residue and fusion-fission production correlates with an increase in the strength of the quasifission reaction channel~\cite{Hinde1995,Hinde1996}.

\begin{table}
  \centering
  \caption{Barriers for average, aligned, and anti-aligned orientations for each measured reaction system.}
  \setlength{\tabcolsep}{5.5pt}
 \setlength{\extrarowheight}{2pt}
\begin{tabular}{ccccccc}
\hline \hline
System &\begin{tabular}{@{}c@{}}V$_{\mathrm{Bass}}$(MeV) \\ (average)\end{tabular}&\begin{tabular}{@{}c@{}}V$_{\mathrm{Bass}}$(MeV) \\ (aligned)\end{tabular}&\begin{tabular}{@{}c@{}}V$_{\mathrm{Bass}}$(MeV) \\ (anti-aligned)\end{tabular} \\
  \hline
  $^{50}$Cr $+ ^{180}$W&196.95&179.84&207.21\\
  $^{50}$Cr $+ ^{186}$W&195.59&180.31&204.70\\
  $^{52}$Cr $+ ^{180}$W&195.75&178.86&205.83\\
  $^{52}$Cr $+ ^{184}$W&194.80&179.00&204.17\\
  $^{54}$Cr $+ ^{180}$W&194.56&177.89&204.51\\
  $^{54}$Cr $+ ^{182}$W&194.12&177.46&204.10\\
  $^{54}$Cr $+ ^{184}$W&193.67&178.05&202.86\\
  $^{54}$Cr $+ ^{186}$W&193.22&178.35&202.10\\
  \hline
  \hline
\end{tabular}
  \label{tab:deformation}
\end{table}

\par
\begin{figure}
\centering
\includegraphics*[width=0.4\textwidth]{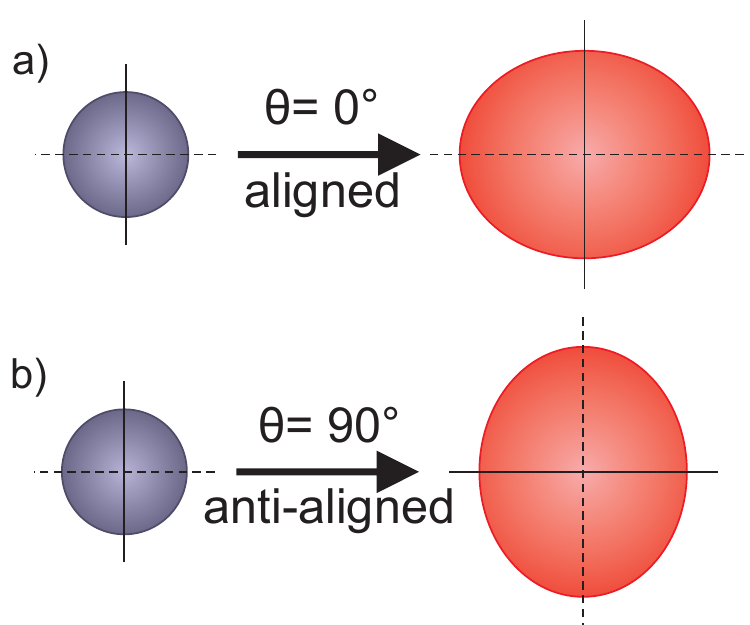}
\caption{Illustration of an aligned (Panel a) and an anti-aligned (Panel b) collision between a spherical projectile and a prolate target.}
\label{fig:collisions}
\end{figure}

It is useful to consider the change in the barrier for the two extreme collision types at an impact parameter b=0.  The nuclear symmetry axes of the two nuclei are aligned and the Cr collides with the tip of the prolate W (Figure ~\ref{fig:collisions}, Panel a) or when the projectile and target nuclear symmetry axes are anti-aligned and the Cr interacts with the elongated side of the prolate W (Figure ~\ref{fig:collisions}, Panel b).

Aligned collisions result in an elongated dinuclear system.  This long shape leads to a preference for quasifission~\cite{Hinde1995, Hinde1996, Hinde2008}.  Conversely, anti-aligned collisions produce a compact dinuclear system that has an enhanced probability of forming a fully fused compound nucleus~\cite{Hinde1995, Hinde1996, Hinde2008}.

\par
The interaction barriers were calculated for the two orientations described above. First, the deformed tungsten nuclei was approximated as an ellipsoid of revolution where the various radii can be calculated from the expression

\begin{equation}
R(\theta,\phi) = R_{\mathrm{avg}}[ 1 + \beta_2 \mathrm{\textbf{Y}}_{20} (\theta, \phi) ]
\end{equation}
where R$_{\mathrm{avg}}$ is the average radius of the two major axes, $\beta_2$ is the deformation parameter along the semi-major axis of interest, and $\mathrm{\textbf{Y}}_{20}$ is a spherical harmonic function ($\mathrm{\textbf{Y}}_{LM}$) where $L$ is 2, $M$ is 0, and $\beta_{4}$ is neglected~\cite{Loveland2006}. In a prolate deformed nucleus, there are two axes of interest: (1) the elongated semi-major axis, along the nuclear symmetry axis which is indicated by the dashed, black line in the example prolate deformed nucleus in Figure~\ref{fig:collisions}, (2) the shortened semi-minor axis indicated by the solid, black line in the example prolate deformed nucleus in Figure~\ref{fig:collisions}.  The limiting case of the semi-major and semi-minor axes can be calculated as:

\begin{eqnarray}
R_{\mathrm{semi-Major}} (\theta, \phi) = R_{\mathrm{avg}}[ 1 - \frac{\beta_2}{4}\sqrt{\frac{5}{\pi}} ]\\
R_{\mathrm{semi-Minor}} (\theta, \phi) = R_{\mathrm{avg}}[ 1 + \frac{\beta_2}{2} \sqrt{\frac{5}{\pi}}  ]
\end{eqnarray}

\begin{figure*}
\centering
\includegraphics*[width=1.0\textwidth]{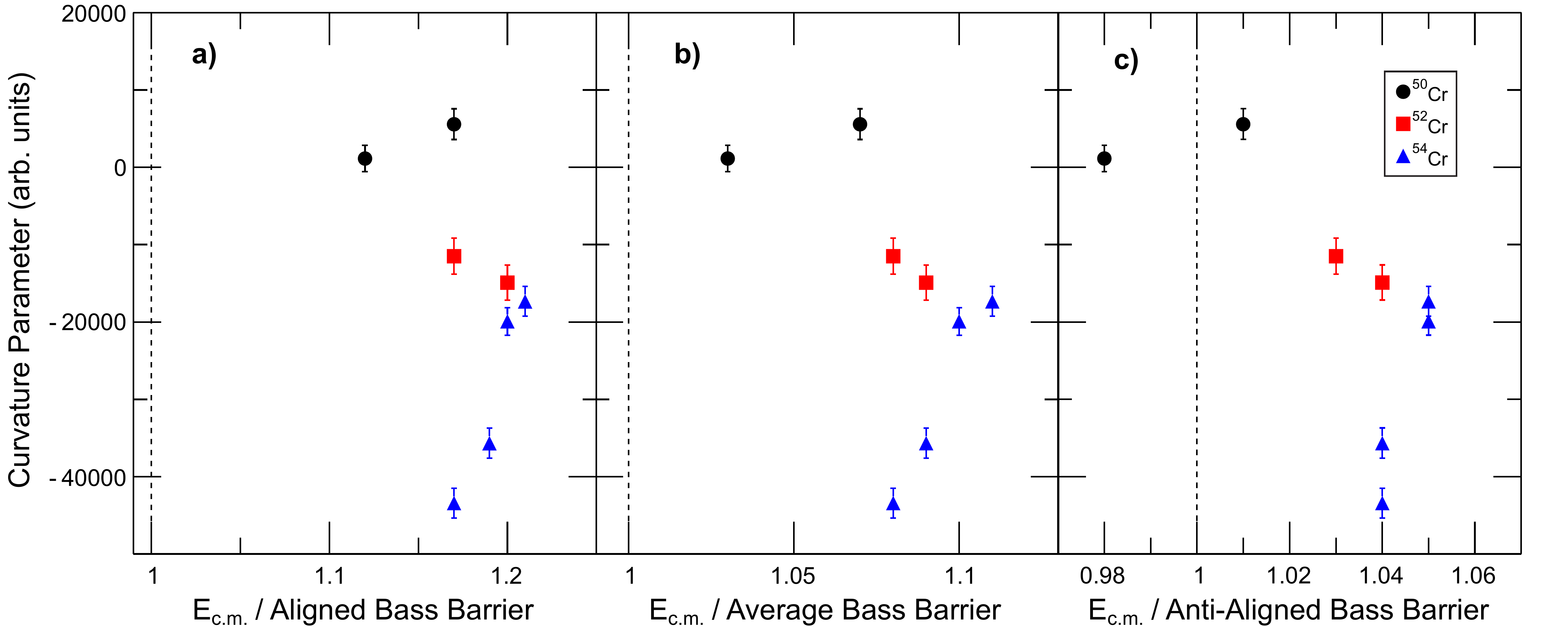}
\caption{(color online) The curvature parameter determined for each system as a function of E$_{\mathrm{c.m.}}$ / V$_{\mathrm{B}}$ for all Cr + W system where V$_{\mathrm{B}}$ is taken as the aligned barrier (Panel a), the average barrier (Panel b),  and the anti-aligned barrier (Panel c).  The dashed line denotes the barrier in each panel.}
\label{fig:CurvVsBarriers}
\end{figure*}

The radius used in the present work was taken to be the Blocki half-density radius~\cite{Bass1977a} given by the expression R$_{\mathrm{avg}}$ = $1.16*\mathrm{A}^{1/3} - 1.39*\mathrm{A}^{-1/3}$. The average, semi-major, and semi-minor radii are listed for the Cr and W nuclei in Table~\ref{tab:radii}. The semi-major and semi-minor axes change by more than 1 fm compared to the average radius, or by about 10$\%$ of the total, because of the strong deformation of the W nuclei. This change in radius has a large effect on the interaction barrier associated with each case considered in this discussion. The interaction radius for a given orientation ($\theta$ as defined in Figure~\ref{fig:collisions}) was determined as R$_{int}$($\theta$) = R$^{\mathrm{Cr}}$($\theta$) + R$^{\mathrm{W}}$ ($\theta$). The Bass barriers~\cite{Bass1977a} for all three orientations (averaged, aligned, and anti-aligned) are shown in Table~\ref{tab:deformation}. As expected the barriers for the aligned collisions are lower than the average, while the barriers for the anti-aligned collisions are higher than the average barriers. In Figure~\ref{fig:CurvVsBarriers}, the curvature is plotted as a function of E$_{\mathrm{c.m.}}$ / V$_{\mathrm{B}}$ for the average (Panel b), aligned (Panel a), and anti-aligned (Panel c) barriers with the dashed vertical line showing the barrier energy in each panel. For the two reactions of $^{50}$Cr, the anti-aligned collision type is strongly hindered as E$_{\mathrm{c.m.}}$ / V$_{\mathrm{B}} (\mathrm{anti-aligned}) = 0.98$ and $1.01$.

\par
Previous work~\cite{Hinde1995, Hinde1996, Hinde2002, Rafiei2008,Thomas2008,Hinde2008,Nishio2001,Mitsuoka2002} has generally shown that quasifission increases near the barrier for heavier systems.  In the present case, fusion in the anti-aligned orientation was seen to be hindered for the $^{50}$Cr systems, therefore, the majority of events that successfully capture and form a dinuclear system do so in the aligned orientation which preferentially leads to the quasifission reaction channel.  The other systems in this work are able to capture in all orientations, including those that preferentially lead to fusion-fission, effectively lowering the curvature of the observed mass distributions and decreasing the amount of quasifission relative to the $^{50}$Cr systems.

\section{Conclusions}
\par
Fission fragment mass distributions were measured for a series of Cr + W reactions at E$^{*}_{\mathrm{CN}} = 52.0$ MeV. A curvature parameter was defined to characterize each mass distribution; a more negative curvature parameter being consistent with the presence of a more prominent fusion-fission reaction channel. The Bohr Independence Hypothesis was not followed for three systems that form the same compound nucleus. Unlike the previously measured Cr~+~W systems, at E$_{\mathrm{c.m.}}$ / V$_{\mathrm{B}} = 1.13$, there was not a strong correlation between the curvature of the distribution and (N/Z)$_{\mathrm{CN}}$. Compound nuclear fissility, mass asymmetry, and rotational energy cannot completely explain the change in quasifission flux between the Cr + W systems. Significant variation in the maximum rotational energy gives an indication as to why the lower energies measured here lead to a higher quasifission flux compared to the previous results for the same Cr + W systems. The strong shift of the fusion barrier depending on the orientation of the deformed W targets significantly hinders the more compact dinuclear orientations that favor fusion, leading to an enhanced quasifission flux for the $^{50}$Cr $+ ^{180}$W and $^{50}$Cr $+ ^{186}$W systems.  In reactions of neutron-rich projectiles and targets to form new heavy and superheavy isotopes it is vital to consider the effect of deformation on quasifission, particularly at energies near to the interaction barrier.
\par

\emph{Acknowledgments.}  The authors are grateful for the high quality beams provided by the staff at the ANU accelerator facility.  This work
is supported by the National Science Foundation under Grants No. PHY-1102511 and No. IIA-1341088, by the U.S. Department of Energy under Grant No.
DE-FG02-96ER40975 with Vanderbilt University, and the Australian Research Council Grants DP160101254, DP140101337, FL110100098, DP130101569, FT120100760, and DE140100784.  This material is based upon work supported by the Department of Energy National Nuclear Security Administration through the Nuclear Science and Security Consortium under Award Number DE-NA0000979.



%

\end{document}